\documentstyle[aps,prl]{revtex}

\begin{document}
\draft
\title{Pinning of slidding collective charge state in a 1D attractive fermion system}
\author{Shu Chen$^1$,Yupeng Wang$^{2,3}$, Fan Yang$^1$,Fu-Cho Pu$^{1,4}$}
\address{$1.$ Institute of Physics and Center for Condensed Matter Physics,
Chinese Academy of Sciences, Beijing, 100080, China\\
$2.$ Cryogenic Laboratory, Chinese Academy of Sciences,
Beijing, 100080, China \\
$3.$ Theoretishe Physik II, Universit\"{a}t Augsburg, 86135, Augsburg, Germany\\
$4.$ Department of Physics, Guangzhou Teacher College,
Guangzhou, 510400, China}
\maketitle

\begin{abstract}
We investigate an interacting fermion model
with boundary potential by using Bethe ansatz method. The ground state
properties of the system and the boundary effect are discussed. It is found
that attractive boundary potential leads to the boundary bound state. An
interesting phenomenon is that the slidding collective charges in a periodic system,
which is formed due to the attractive interaction among the fermions, will be pinned around the
boundary, as long as the negative boundary potential is strong enough.
\end{abstract}
\section{Introduction}
There has been extensive interest in the investigation of low-dimensional
 correlated fermion systems in the recent years. It is well known that the
perturbation techniques are not very effective tools in dealing 
 with the one-dimensional systems. The 1D
systems, although they are somewhat artificial, can give some valuable
information on the role of correlation effects in higher dimensions and thus
are theoretically meaningful. Many approaches, such as bosonization and
renormalization group techniques, have been successfully applied in this
field and some fruitful results are obtained. Usually, studies on the
exactly solvable models can provide an exactly theoretical
understanding to the more complicated interacting systems. Thus the quantum
integrable models, which have a long history in itself, again attract much
attention.

Some recent works\cite{Wang,Wang2} show that the behavior of a
single impurity in the 1D quantum systems is rather different from that in a
Fermi liquid. Kane and Fisher\cite{Kane} first investigated a 1D repulsive
interacting system in the presence of a potential barrier and pointed out
that it corresponds to a chain disconnected at the barrier site at low
energy scales. This can be effectively described by the open boundary
condition and is well investigated by the boundary conformal field theory
\cite{Affleck}. The open boundary problem is also studied by using the Bethe
ansatz method\cite{Gaudin,Schulz,Skly}. Attempting to understand the
effect of boundary potential to an interacting fermion system, we study a
 spinless interacting model with boundary potential by the Bethe
ansatz method. The results show that there exists a critical value for the
attractive boundary potential. When it goes through the critical value, the strong
attractive boundary potential leads to a boundary bound state. Moreover, the
attractive interaction among the fermions also has a critical value. When
the interaction is strong enough, a collective slidding charge state forms.
The attractive
boundary potential will make the whole collective state to be pinned down
around the boundary barrier site. This phenomenon provides us a
meaningful example  how a local potential influences the global
properties in a strongly correlated fermion system.
\section{The model and Bethe ansatz}

Consider a 1D lattice with $L$ sites and N particles. The Hamiltonian
reads
\begin{equation}
\begin{array}{l}
H=\sum_{j=1}^{L-1}[-t(a_j^{+}a_{j+1}+a_{j+1}^{+}a_j)+  
Un_jn_{j+1}]+p_1n_1+p_Ln_L,
\end{array}
\end{equation}
where $a_j(a_j^{+})$ are fermion annihilation (creation) operators and $t,$ $
U,$ $p_1,$ $p_L$ are hopping amplitude, coupling and boundary potential
constants respectively. The hopping parameter is defined to be positive $
(t>0)$. As a matter of fact, this model is equivalent to the $XXZ$
Heisenburg model\cite{Yang,Alc}by a Jordan-Wigner transformation\cite
{Pu}. For simplicity, we take $p_L=\infty ,$ which means that the particles
arriving at the right boundary are completely reflected by a infinite high
wall, and only consider the effect of the left boundary. To make the discussion clearer,
we rederive some known results about this model in this section.

Taking the eigenstate as 
\[
\mid \Psi \rangle _N=\sum\limits_{x_1\cdots x_N}\varphi (x_1\cdots
x_N)a_{x_1}^{+}\cdots a_{x_N}^{+}\mid 0\rangle, 
\]
we get the eigenequation 
\begin{equation}
\begin{array}{l}
-t\sum\limits_k\varphi (x_1...x_k\pm 1,...x_N)+  
U\sum\limits_{k<l}\delta _{x_l+1,x_k}\varphi (x_1\cdots
x_N)+[p_1\sum\limits_k\delta (x_k-1) \\ 
+p_N\sum\limits_k\delta (x_k-L)]\varphi (x_1\cdots x_N)=Eg(x_1\cdots x_N).
\end{array}
\end{equation}
Considering the two-particle state, in the region of $x_1<x_2,$ we take the
wavefunction as Bethe type 
\begin{equation}
\begin{array}{l}
\varphi \left( x_1,x_2\right) =\sum_{r1,r2}[  
A_{12}\left( r_1,r_2\right) \exp (ir_1k_1x_1+ir_2k_2x_2)+  
A_{21}\left( r_2,r_1\right) \exp (ir_2k_2x_1+ir_1k_2x_2)],
\end{array}
\end{equation}
where $r_{1,2}=\pm 1$ indicates that the particles move toward  right or
 left$.$ Far away from the boundary $(2<x_1,x_2<L),$ when the two
particles are not neighboring $(x_1\neq x_2-1)$, it is easy to get the
eigenvalue from the eigenequation 
\begin{equation}
E=-2t\left( \cos k_1+\cos k_2\right) .
\end{equation}
When the two particles are neighboring $(x_1=x_2-1)$, solving the eigenvalue
equation, it is readily to obtain the following relation 
\begin{equation}
A_{21}\left( r_2,r_1\right) =S(r_1k_1,r_2k_2)A_{12}\left( r_1,r_2\right)
\end{equation}
with 
\begin{equation}
\begin{array}{l}
S(r_1k_1,r_2k_2)=-e^{i\theta (r_1k_1,r_{2k_2})}=-\frac{1+\exp i(r_1k_1+r_2k_2)+U/t\exp ir_2k_2}{1+\exp
i(r_1k_1+r_2k_2)+U/t\exp ir_1k_1}
\end{array}
\end{equation}

Comparing to the periodic boundary case, the wavefunction with open boundary
includes some coefficients  corresponding to backscattering $
\left( r=-1\right) $ waves. In addition, the boundary conditions
give some limitation to these coefficients. When a particle is located at
the left boundary{\bf \ }$\left( x_1=1\right) ,$ the eigenvalue equation
implies the relation $A_{12}\left( -,r_2\right) =S_L(k_1)A_{12}\left(
+,r_2\right) $ or $A_{21}\left( -,r_1\right) =S_L(k_2)A_{21}\left(
+,r_1\right) $ with 
\begin{equation}
S_L(k_i)=-\frac{e^{-ik_i}+p_1/t}{e^{ik_i}+p_1/t}e^{i2k_i}.
\end{equation}
For a particle at the right boundary{\bf \ }$\left( x_2=L\right)
, $ the eigenvalue equation also implies the relation $A_{12}\left(
r_1,-\right) =S_R(k_2)A_{12}\left( r_1,+\right) $ or $A_{21}\left(
r_2,-\right) =S_R(k_1)A_{21}\left( r_2,+\right) ,$ with 
\begin{equation}
S_R(k_i)=-e^{i2k_iL}.
\end{equation}

In general, we suppose the wavefunction has the following form 
\begin{eqnarray}
\varphi \left( x_1,\cdots ,x_N\right)=\sum_{p,Q,r_p}(-1)^{I(Q)}A_P(r_p) 
\prod_{i=1}^ne^{ir_{p_i}k_{p_i}x_{Q_i}}\theta (x_{Q_1} <\cdots <x_{Q_N}),
\end{eqnarray}
where $I(Q)$ is the parity of the permutation of $Q$. With the same
procedure, a series of relations can be obtained
by using the eigenequation $(2).$ Similar to the two particle case, the
eigenvalue is given by 
\begin{equation}
E=-2t\sum_{j=1}^N\cos k_{_j}.
\end{equation}
When two particles are neighboring, we get 
\begin{eqnarray}
 A_{p_{1\cdots }p_{j+1,}p_{j\cdots }p_N}\left( r_1,\cdots
r_{p_{j+1}},r_{p_j},\cdots r_{p_N}\right)
=S(r_{p_j}k_{p_j},r_{p_{J+1}}k_{p_{J+1}}) A_{p_{1\cdots }p_{j,}p_{j+1\cdots }p_N}\left( r_1,\cdots
r_{p_j},r_{p_{j+1}},\cdots r_{p_N}\right) .  
\end{eqnarray}
with $S(k_{p_j},k_{p_{j+1}})$ given by $\left( 6\right) $. When the particle
is at the boundaries, the eigen equation gives 
\begin{eqnarray}
A_p\left( -,\cdots \right) =S_L(k_{p_1})A_p\left( +,\cdots \right) ,{~~~~~~}
A_p\left( \cdots ,-\right) =S_R(k_{p_N})A_p\left( \cdots ,+\right) ,
\end{eqnarray}
with $S_L(k_i),$ $S_R(k_i)$ given by $\left( 7\right) $ and $\left( 8\right)
.$ From the relations (10-12) we conclude 
\begin{eqnarray*}
A_{p_{1\cdots }p_N} =S_L^{-1}(k_{p_1})S(k_{p_2},-k_{p_1})\cdots
 S(k_{p_N},-k_{p_1})S_R(k_{p_1})S(k_{p_1},k_{p_N}) 
 \cdots S(k_{p_1},k_{p_2})A_{p_1p_2\cdots p_N}.
\end{eqnarray*}
For simplicity, we use $A_{p_1\cdots p_j\cdots p_N}$ to represent $%
A_{p_{1\cdots }p_{j\cdots }p_N}\left( +,\cdots ,+,\cdots +\right) $ and $%
A_{p_1\cdots \overline{p_j}\cdots p_N}$ to represent $A_{p_{1\cdots
}p_{j\cdots }p_N}\left( +,\cdots ,-,\cdots +\right) $ in the following text$%
. $ By taking $p_1=j,$ we get 
\begin{equation}
S_L(-k_j)S_R(k_j)\prod\limits_{i=1(\neq j)}^Ne^{i\theta
(k_i,-k_j)}e^{i\theta (k_j,k_i)}=1.
\end{equation}
Representing $%
S_L(k_i)=-e^{i\theta _L(k_i)}e^{i2k_i}$ and $S_R(k_i)=-e^{i2k_iL},$ we
have 
\begin{equation}
e^{-ik_j2(L-1)}=\prod\limits_{i=1(\neq j)}^ne^{i\theta _L(-k_j)}e^{i\theta
(k_j,k_i)}e^{i\theta (k_i,-k_j)}.
\end{equation}

For convenience, we take $t=1$ and put $\Delta =\frac U{2t}$ in the
following discussion. $\Delta >0$ or $\Delta <0$ corresponds to the
repulsive or attractive interaction respectively$.$ As is well known, $
\Delta  =\pm1$ are two  critical points. $\mid \Delta \mid >1$ indicates
the strong coupling regions while $\mid \Delta \mid <1$ indicates
weak coupling region. In the next section we discuss how the boundary potential influences
the properties of the ground state of the whole system in different parameter
regions.

\section{Boundary bound states}
We discuss first the gapless critical region. For{\bf \ } $\mid
\Delta \mid <1,$ it's convenient to represent{\bf \ }$\Delta =\cos \eta $ ($%
U=2\cos \eta $) with $\eta \in (0,\pi )${\bf .\ }The repulsive or attractive
interaction is decided by $\eta \in (0,\pi /2)$ or $(\pi /2,\pi )$
respectively. We parameterize $k_j$ as $k_j=\phi (\lambda _j,\frac \eta 
2)$ with $\phi (a,b)$ defined by $\phi (a,b)=2\arctan (\tanh a\cot b).$ It's
readily to get $\theta (k_j,k_i)=\phi (\lambda _i-\lambda _j,\eta ).$ Put 
\begin{equation}
p_1=\frac{e^{i\Gamma +i\eta }-1}{e^{i\Gamma }-e^{i\eta }}=\frac{\sin \frac 12%
(\Gamma +\eta )}{\sin \frac 12(\Gamma -\eta )}.
\end{equation}
It follows that $e^{i\theta _L(k_j)+ik_j}=-\sinh (\lambda _j-\frac{
i\Gamma }2)/\sinh (\lambda _j+\frac{i\Gamma }2)$. The Bethe ansatz
equation is thus reduced to 
\begin{eqnarray}
\left( \frac{\sinh (\lambda _j+\frac i2\eta )}{\sinh (\lambda _j-\frac i2
\eta )}\right) ^{2L-1}\frac{\sinh (\lambda _j-\frac{i\Gamma }2)}{\sinh
(\lambda _j+\frac{i\Gamma }2)} 
=\prod\limits_{r=\pm 1}\prod\limits_{i=1(\neq j)}^N\frac{\sinh (\lambda
_j-r\lambda _i+i\eta )}{\sinh (\lambda _j-r\lambda _i-i\eta )},
\end{eqnarray}
and the eigenenergy is 
\begin{eqnarray}
E=2t\sum_{j=1}^N[\cos \eta -\frac{\sin ^2\eta }{\cosh 2\lambda _j-\cos \eta 
}]
\end{eqnarray}
It should be notified that the Bethe ansatz equation with open boundary
condition is reflecting invariant, which means $\lambda
_j$ and $-\lambda _j$ corresponds to the same state, so we only
need to choose one in dealing with concrete problem.

From equation $\left( 16\right) ,$ we observe that $\lambda _j=i\frac \Gamma 
2$ $\left( \text{or }-i\frac \Gamma 2\right) $ is a possible solution of
the Bethe ansatz equation in the condition of $\mid \frac{\sin
\frac 12(\Gamma +\eta )}{\sin\frac 12(\Gamma -\eta )}\mid $ $>1$ $\left( \mid p_1\mid >1\right) $when $%
L\rightarrow \infty .$ It can also be seen from $\left( 17\right) ,$ when $%
\cos \Gamma >\cos \eta $ $,$ the energy contributed by the boundary mode
is lower than that of a real mode. For example $\eta \in \left( 0,\pi /2\right) ,$
taking $0<\frac 12(\Gamma +\eta )<\frac \pi 2,$ and $-\frac \pi 2<\frac 12%
(\Gamma -\eta )<0,$ the condition $\mid p_1\mid $ $>1$ is satisfied. By the
limitation $0<\Gamma <\frac \pi 2$ $,$ the condition $\cos \Gamma >\cos \eta 
$ is satisfied if $\Gamma <\eta $. This implies $p_1<-1.$

So far, we have learned that repulsive or the small attractive $\left(
-1<p_1<0\right) $ potential does not produce the boundary bound state. The
BA equation has no imaginary solution in the ground state. The effect of
boundary potential of this case has been extensively investigated by many
authors\cite{Alc}. The strong attractive $\left( p_1<-1\right) $
potential produces a boundary bound state and the corresponding energy
contributed by the boundary bound state is 
\[
e_b=2t[\cos \eta -\frac{\sin ^2\eta }{\cos \Gamma -\cos \eta }], 
\]
with $0<\Gamma <\eta <\frac \pi 2.$ In this case, the solutions of BA
equation to the ground state consist of a boundary imaginary mode and $N-1$ real
mode. This problem has also been discussed by many authors\cite{Skorik,Skorik2,Bed}
for $XXZ$ and Hubbard model. 

For{\bf \ }$\triangle =-1,$ the interaction term is attractive. In this
case, the model corresponds to the well known $XXX$ ferromagnetic spin chain
\cite{Bethe} with fixed magnetization. Parameterizing $k_j=2%
\mathop{\rm arccot}
2\lambda _j,$ we have $\theta (k_j,k_i)=2\arctan (\lambda _j-\lambda _i).$
Putting 
\begin{equation}
p_1=\frac{1+\Gamma }{1-\Gamma },
\end{equation}
the Bethe ansatz equation is given by 
\begin{eqnarray}
 \left( \frac{\lambda _j+\frac i2}{\lambda _j-\frac i2}\right) ^{2L-1}
\frac{\lambda _j-\frac i2\Gamma }{\lambda _j+\frac i2\Gamma }  
=\prod\limits_{r=\pm 1}\prod\limits_{i\neq j}^N\frac{\lambda
_j-r\lambda _i+i}{\lambda _j-r\lambda _i-i}.
\end{eqnarray}
The eigenvalue is given by 
\begin{equation}
E=t\sum_{j=1}^N(\frac 4{4\lambda _j^2+1}-2).
\end{equation}

With periodic boundary condition, all possible  string solutions
of the Bethe ansatz equation are $\lambda _{\alpha ,k}^n=\lambda _\alpha ^n-
\frac i2(n+1-2k)$ with $k=1,\cdots ,n.$ Notice $\frac 1{\lambda ^2+1/4}=i[
\frac 1{\lambda +i/2}-\frac 1{\lambda -i/2}].$ Thus a $n$-string carries
the energy 
\[
\sum_{k=1}^n\frac 4{4\lambda _{\alpha ,k}^n{}^2+1}=\frac n{\lambda _\alpha
^n{}^2+n^2/4}. 
\]
Representing $M_n$ as the number of $n$-strings, the relation $
N=\sum_{n=1}^\infty nM_n$ holds. It can be proved 
\begin{equation}
\frac N{\lambda _j^2+N^2/4}<\sum_{n=1}^{N-1}M_n\frac n{\lambda _j^2+n^2/4}.
\end{equation}
Thus the $N$ string state is favorable since it contributes much lower
energy than the other states. The corresponding ground state wavefunction
describes a bound state which decays exponentially as a function of
coordinate differences $\mid x_j-x_i\mid $ with $j\neq i.$

It can be seen from $\left( 19\right) ,$ $\lambda _j=\frac i2\Gamma $ is a
solution of Bethe equations in the limit of $L\rightarrow \infty $ if $|p_1|>1$. 
Owing to $\lambda _j=\frac i2\Gamma $
is a solution of Bethe equation,  the following type of
strings
\begin{equation}
\lambda _k^n=\frac i2(\Gamma +2k-2)
\end{equation}
with $k=1,\cdots ,n$ are also possible solutions$.$ The energy of an
imaginary mode is  
\[
e_k^n=\frac 4{4(\lambda _k^n)^2+1}=\frac 4{1-(\Gamma +2k-2)^2}. 
\]
Now, it's easy to see that when $\Gamma >1$
the energy contributed by the boundary string is lower than that of a real mode.

As discussed, the attractive boundary potential produces a boundary
bound state as soon as it is stronger than the hopping amplitude. In the
case of attractive interaction, the ground state of systems will be a
boundary string state with length $N$ in the form $\left( 23\right) .$ In
virtue of the relation 
\[
2\sum_{k=1}^n\frac 1{1-(\Gamma +2k-2)^2}=\frac 1{\Gamma +2n-1}-\frac 1{%
\Gamma -1}, 
\]
the ground state energy is given by 
\begin{eqnarray}
E=t\sum_{k=1}^N(\frac 4{4(\lambda _k^N)^2+1}-2) 
=2t(\frac 1{\Gamma +2N-1}-\frac 1{\Gamma -1})-2Nt.
\end{eqnarray}
So far, we have shown all the particles are bounded by the attractive
boundary potential in the ground state. The wavefunction decays exponentially
 away from the left boundary. In order to show it explicitly, we
give an example of two particle case in the appendix.
Generally, for a $n$ particle system, the ground state is a  boundary $n$-strings $\left( 22\right) $,
the corresponding wavefunction is given by 
\begin{equation}
\varphi \left( x_1,\cdots x_n\right) =cf\left( \lambda _1^n\right)
^{-x_1}f\left( \lambda _2^n\right) ^{-x_2}\cdots f\left( \lambda _n^n\right)
^{-x_3},
\end{equation}
with $f(\lambda _j)=e^{ik_j}=%
\frac{\lambda _j+\frac i2}{\lambda _j-\frac i2}.$ Therefore, with the presence of the attractive
impurity potential, the slidding collective charges will be pined and loses its mobility. 
By the expression of energy $\left( 24\right) ,$ we learn that the
stronger the attractive potential is, the more quickly the 
wavefunction decays away from the boundary.

For the repulsive interaction $U=2$ $\left( \Delta =1\right) ,$ we
re-parameterize $k_j=2\arctan 2\lambda _j,$ thus $\theta (k_j,k_i)=2\arctan
(\lambda _i-\lambda _j).$ Putting $p_1=\frac{\Gamma +1}{\Gamma -1},$ the
Bethe ansatz equation is also given by $\left( 19\right) ,$ but the
eigenvalue is given by 
\begin{equation}
E=t\sum_{j=1}^N(2-\frac 4{4\lambda _j^2+1}).
\end{equation}
Similar to the case $\Delta =-1$, the Bethe ansatz equation has the
imaginary solution $\lambda _j=i\frac \Gamma 2$ $\left( \text{or }-i\frac 
\Gamma 2\right) $ if $\mid p_1\mid >1.$ Thus $\lambda _k^n=\frac i2(\Gamma
+2k-2)$ is also the possible solution of BA equation. From $\left( 25\right)
,$ we see if $\Gamma \in (0,1)$ the energy of the imaginary mode $%
\lambda ^1=i\frac \Gamma 2$ is lower than those of the real modes. However, the other
imaginary solutions for $n>1$ are not favorable in energy and thus corresponds
to the highly excited state. In this case, the ground state is composed of
an imaginary mode $\lambda =i\frac \Gamma 2$ with $\Gamma \in (0,1)$ and $%
N-1 $ real mode.

In the following, we  discuss the ground state properties of 
strong interaction cases, $\mid \bigtriangleup \mid >1.$
First, we consider the attractive interaction case. Owing to $%
\bigtriangleup <-1,$ we can parameterize it as $\Delta =-\cosh \eta $ with $%
\eta >0${\bf .} By putting   
\begin{equation}
k_j=2%
\mathop{\rm arccot}
(\tan \lambda _j\coth \frac 12\eta ),
\end{equation}
\begin{equation}
\theta (k_j,k_i)=2\arctan [\tan (\lambda _j-\lambda _i)\coth \eta ],
\end{equation}
\begin{equation}
p_1=-\frac{e^{\Gamma +\eta }-1}{e^\Gamma -e^\eta }=-\frac{\sinh \frac 12%
(\Gamma +\eta )}{\sinh \frac 12(\Gamma -\eta )},
\end{equation}
the Bethe ansatz equation can be reduced to 
\begin{eqnarray}
\left( \frac{\sin (\lambda _j+\frac 12i\eta )}{\sin (\lambda _j-\frac 12
i\eta )}\right) ^{2L-1}\frac{\sin (\lambda _j-i\frac \Gamma 2)}{\sin
(\lambda _j+i\frac \Gamma 2)} 
=\prod\limits_{r=\pm 1}\prod_{i\neq j}^N\frac{\sin (\lambda _j-r\lambda
_i+i\eta )}{\sin (\lambda _j-r\lambda _i-i\eta )}.
\end{eqnarray}
The eigenvalue is given by 
\begin{equation}
E=2t\sum_{j=1}^N[\frac{\sinh ^2\eta }{\cosh \eta -\cos 2\lambda _j}-\cosh
\eta ]
\end{equation}

In the attractive interaction case ($\Delta =-\cosh \eta $), the energy
contributed by a string is lower than those of the real modes. In the period
boundary condition, all the possible types of string solutions are 
\begin{equation}
\lambda _{\alpha ,k}^n=\lambda _\alpha ^n-\frac i2\eta (n+1-2k)
\end{equation}
with $k=1,\cdots ,n.$  The energy of an n-string
is given by 
\begin{equation}
e_n(\lambda )=\sinh \eta \frac{\sinh n\eta }{\cosh n\eta -\cos 2\lambda }%
-n\cosh \eta .
\end{equation}
It can be proved that 
\[
\frac{\sinh N\eta }{\cosh N\eta -\cos 2\lambda }<\sum_{n=1}^{N-1}M_n\frac{%
\sinh n\eta }{\cosh n\eta -\cos 2\lambda }. 
\]
Under the periodic boundary condition, the ground state in strong
attractive case is given by an N-string \cite{Albert,Gaudin2}. Similar to the case $\triangle
=-1, $ the ground state wavefunction decays exponentially 
as a function of coordinate differences $\mid
x_j-x_i\mid $ for all $j\neq i.$

The Bethe ansatz equation $\left( 29\right) $ has an imaginary solution $%
\lambda _j=i\frac \Gamma 2$ for $\Gamma >0$ in the limit $L\rightarrow
\infty $. This solution corresponds to the boundary bound state$.$ The
imaginary mode contributes the energy 
\begin{equation}
e_b(i\frac \Gamma 2)=\frac{\sinh ^2\eta }{\cosh \eta -\cosh \Gamma }-\cosh
\eta .
\end{equation}
For $\Gamma >\eta ,$ the imaginary mode $i\frac \Gamma 2$ contributes
much lower energy than that of a real mode. 
Similar to the discussion for $\Delta =-1$ case, if $i\frac \Gamma 2$ is a
solution of $BA$ equation, the following  string  
\begin{equation}
\lambda _k^n=\frac i2[\Gamma +2(k-1)\eta ]
\end{equation}
with $k=1,\cdots ,n,$ are also solutions. The energy of an imaginary mode is written as 
\[
e_k^n=\frac{\sinh ^2\eta }{\cosh \eta -\cosh [\Gamma +2(k-1)\eta ]}-\cosh
\eta . 
\]
Now, it's easy to see that the energy
of this state is lower than thoes of any other states. In this case, the ground
state solution is not a slidding $N$-string as $\left( 31\right) $ but a boundary $N$-string 
given by $\left( 34\right) $. 
Put $\Gamma =\Gamma ^{^{\prime }}-(N-1)\eta .$ In virtue of 
\begin{eqnarray*}
\sum_{k=1}^n\frac{\sinh ^2\eta }{\cosh \eta -\cosh [\Gamma ^{^{\prime
}}+(2k-n-1)\eta ]} 
=\sinh \eta \frac{\sinh n\eta }{\cosh n\eta -\cosh \Gamma ^{^{\prime }}},
\end{eqnarray*}
 we obtain the ground state energy as
\begin{eqnarray*}
E =2t\sum_{k=1}^Ne_k^N 
=\frac{2t\sinh \eta \sinh N\eta }{\cosh N\eta -\cosh [\Gamma +(N-1)\eta ]}
-2tN\cosh \eta .
\end{eqnarray*}

Similar to the $\bigtriangleup =-1$ case, the ground state wavefunction is
given by 
\begin{equation}
\varphi \left( x_1,\cdots x_n\right) =cf\left( \lambda _1^n\right)
^{-x_1}f\left( \lambda _2^n\right) ^{-x_2}\cdots f\left( \lambda _n^n\right)
^{-x_3},
\end{equation}
with 
\[
f\left( \lambda _k^n\right) =\frac{\sinh \frac 12[\Gamma +(2k-1)\eta ]}{%
\sinh \frac 12[\Gamma +(2k-3)\eta ]}. 
\]

For{\bf \ }$\Delta =\cosh \eta $ $(\eta >0),$ we introduce the notations
\begin{equation}
k_j=2\arctan (\tan \lambda _j\coth \frac 12\eta ),
\end{equation}
\begin{equation}
\theta (k_j,k_i)=2\arctan [\tan (\lambda _i-\lambda _j)\coth \eta ],
\end{equation}
\begin{equation}
p_1/t=\frac{e^{\Gamma +\eta }-1}{e^\Gamma -e^\eta }=\frac{\sinh \frac 12%
(\Gamma +\eta )}{\sinh \frac 12(\Gamma -\eta )}.
\end{equation}
The eigenvalue is expressed as
\begin{eqnarray}
E=2t\sum_{j=1}^N[\cosh \eta -\frac{\sinh ^2\eta }{\cosh \eta -\cos 2\lambda
_j}]
\end{eqnarray}

In this case, 
the string solutions correspond to highly excited states$.$ Now, we can see $%
\lambda _j=i\frac \Gamma 2$ is a possible solution of BA equation $\left(
29\right) $ if the condition $\mid p_1\mid >1$ is satisfied. It is easy to
see that the energy of the imaginary mode $\lambda _j=i\frac \Gamma 2$ is
smaller than those of real modes when $0<\Gamma <\eta $ $\left(
p_1<-1\right) $ for the repulsive interaction $\left( \Delta >1\right) .$
\section{Conclusion}
In summary, we study an interacting spinless fermion system with boundary
potential. By analyzing the Bethe ansatz equation and the eigenenergy, we
discuss the boundary effect of the ground state. Our results show that, when
the attractive potential increases to the value of the hopping amplitude, there
will be a boundary bound state which leads to a fermion to be bounded on the
first site. When the attractive potential is weaker comparing to the hopping
amplitude, the attractive potential is not enough to resist the hopping
charges. For the attractive
interaction $U<0$, there exists a critical value $U=-2t$ $(\bigtriangleup
=-1).$ When the attractive interaction exceeds the critical value, it
makes all particles correlate with each other in a form similar to the
''Cooper's pair''. The corresponding ground state wavefunction decays
exponentially as a function of coordinate differences $\mid x_j-x_i\mid $
for all $j\neq i.$ In this case, the boundary bound state will influence the
properties of the whole system drastically. Owing to the strong correlation
effect among the fermions, the collective charge composite is pinned down in the
neighborhood of the boundary barrier site when the attractive potential is
strong. This is much different from the common Fermi liquid
picture, in which a single potential barrier can not change
 the global properties of a system. This exactly solved model, although is very
simple, reveals a meaningful picture of a strongly correlated fermion system.

\appendix{\bf Appendix}

For the two particles system, the wavefunction is given by $\left( 3\right) $%
. We note A$_{21}\left( r_2,r_1\right) =S(r_1k_1,r_2k_2)A_{12}\left(
r_1,r_2\right) $ with 
\[
S(r_1k_1,r_2k_2)=\frac{r_1\lambda _1-r_2\lambda _2-i}{r_1\lambda
_1-r_2\lambda _2+i}. 
\]
For convenience, we represent 
\[
f(\lambda _j)=e^{ik_j}=\frac{\lambda _j+\frac i2}{\lambda _j-\frac i2}, 
\]
thus the wavefunction can be expressed as 
\begin{eqnarray}
\varphi \left( x_1,x_2\right)=\sum_{r1,r2}A_{12}\left( r_1,r_2\right)
[f\left( \lambda _1\right) ^{r_1x_1}f\left( \lambda _2\right) ^{r_2x_2} 
+S(r_1k_1,r_2k_2)f\left( \lambda _2\right) ^{r_2x_1}f\left( \lambda
_1\right) ^{r_1x_2}].
\end{eqnarray}
Using the relation $\left( 12\right) $ and $\left( 13\right) ,$ we have $A_{%
\overline{1}2}=S_L^1A_{12},$ $A_{1\overline{2}}=S_R^2A_{12},$ $A_{\overline{1%
}\overline{2}}=S_L(\lambda _1)S_R(\lambda _2)A_{12}.$ 
We note 
\[
S_L(\lambda _1)=-\frac{(\lambda _1+\frac i2)(\lambda _1+\frac i2\Gamma )}{%
(\lambda _1-\frac i2)(\lambda _1-\frac i2\Gamma )}. 
\]

Using the Bethe ansatz equation $\left( 14\right) ,$ we have$.$ 
\[
S_R(k_2)=-e^{i2k_2L}=S_L(\lambda _2)S_{12}S_{\overline{2}1}. 
\]
Thus the wavefunction is written as 
\begin{equation}
\begin{array}{l}
\varphi \left( x_1,x_2\right) =  
A_{12}[f\left( \lambda _1\right) ^{x_1}f\left( \lambda _2\right)
^{x_2}+S_{12}f\left( \lambda _2\right) ^{x_1}f\left( \lambda _1\right)
^{x_2}]+  
A_{\overline{1}2}[f\left( \lambda _1\right) ^{-x_1}f\left( \lambda _2\right)
^{x_2}+S_{\overline{1}2}f\left( \lambda _2\right) ^{x_1}f\left( \lambda
_1\right) ^{-x_2}]+ \\ 
A_{1\overline{2}}[f\left( \lambda _1\right) ^{x_1}f\left( \lambda _2\right)
^{-x_2}+S_{1\overline{2}}f\left( \lambda _2\right) ^{-x_1}f\left( \lambda
_1\right) ^{x_2}]+ 
\ A_{\overline{1}\overline{2}}[f\left( \lambda _1\right) ^{-x_1}f\left(
\lambda _2\right) ^{-x_2}+S_{\overline{12}}f\left( \lambda _2\right)
^{-x_1}f\left( \lambda _1\right) ^{-x_2}],
\end{array}
\end{equation}
with $A_{\overline{1}2}=S_L(\lambda _1)A_{12},$ $A_{1\overline{2}%
}=S_L(\lambda _2)S_{12}S_{\overline{2}1}A_{12},$ $A_{\overline{1}\overline{2}%
}=S_L(\lambda _1)S_L(\lambda _2)S_{12}S_{\overline{2}1}A_{12}.$ If we take 
$\lambda _1=\frac i2\Gamma ,$ $\lambda _2=\frac i2(\Gamma
+2),$ then $S_{12}=\infty ,$ $S_L(\lambda _1)=\infty $. That means only one term does not
disappear for $\lambda _2-\lambda _1=i.$ The wavefunction has the form 
\[
\varphi \left( x_1,x_2\right) =cf\left( \lambda _1\right) ^{-x_1}f\left(
\lambda _2\right) ^{-x_2} 
\]
If we take $\lambda _1=-\frac i2\Gamma ,$ $\lambda _2=-\frac i2(\Gamma +2),$ 
$S_{12}=0,$ $S_L(\lambda _1)=0$, the wave function is written as 
\[
\varphi \left( x_1,x_2\right) =cf\left( \lambda _1\right) ^{x_1}f\left(
\lambda _2\right) ^{x_2}. 
\]
The wavefunction can also be expressed as 
\begin{eqnarray*}
\varphi \left( x_1,x_2\right) =\sum_{r1,r2}A_{21}\left( r_2,r_1\right)
[f\left( \lambda _2\right) ^{r_2x_1}f\left( \lambda _1\right) ^{r_1x_2} 
+S^{-1}(r_1k_1,r_2k_2)\ \ f\left( \lambda _1\right) ^{r_1x_1}f\left(
\lambda _2\right) ^{r_2x_2}].
\end{eqnarray*}
It can be written as the similar form of $\left( 41\right) $ with $%
A_{\overline{2}1}=S_L(\lambda _2)A_{21},$ $A_{2\overline{1}}=S_{\overline{1}%
2}S_L(\lambda _1)S_{12}^{-1}A_{21},$ $A_{\overline{2}\overline{1}%
}=S_L(\lambda _1)S_L(\lambda _2)S_{\overline{2}1}A_{21}.$ If we take $%
\lambda _2=\frac i2\Gamma ,$ $\lambda _1=\frac i2(\Gamma +2),$ $%
S_{21}=\infty $, $S_L(\lambda _2)=\infty ,$ it is reduced to 
\[
\varphi \left( x_1,x_2\right) =cf\left( \lambda _2\right) ^{-x_1}f\left(
\lambda _1\right) ^{-x_2}. 
\]
If we take $\lambda _2=-\frac i2\Gamma ,$ $\lambda _1=-\frac i2(\Gamma +2),$ 
$S_{21}=0$, $S_L(\lambda _2)=0,$the wave function is written as 
\[
\varphi \left( x_1,x_2\right) =cf\left( \lambda _2\right) ^{x_1}f\left(
\lambda _1\right) ^{x_2}. 
\]
So far, we can see, the wavefunction can be always represented as the form 
\begin{eqnarray*}
\varphi \left( x_1,x_2\right) &=&cf\left( \lambda _1^2\right) ^{-x_1}f\left(
\lambda _2^2\right) ^{-x_2} \\
\ &=&c\left( \frac{\Gamma +1}{\Gamma -1}\right) ^{-x_1}\left( \frac{\Gamma +3%
}{\Gamma +1}\right) ^{-x_2}.
\end{eqnarray*}
{\bf Acknowledgment}

One of the authors (Y Wang) acknowledges the financial supports from Alexander von
Humboldt-Stiftung and CNFNS.

\end{document}